\documentclass[aip, jcp, notitlepage, floatfix, reprint, citeautoscript, twocolumn]{revtex4-1}

\usepackage{amsmath}
\usepackage{amsfonts}
\usepackage{graphicx}
\usepackage{dcolumn}
\usepackage[table]{xcolor}
\usepackage[version=3]{mhchem}
\usepackage{transparent}
\usepackage{gensymb}
\usepackage{tabularx}
\usepackage{etoolbox}
\usepackage[hidelinks]{hyperref}
\usepackage{bm}
\usepackage{relsize}
\usepackage[normalem]{ulem}

\def\maxfloatwidth{%
  \ifdim\columnwidth>246.0pt
  300.0pt  \else
  \columnwidth
  \fi
}

\newcommand{\mrm}[1]{\mathrm{#1}}
\newcommand{\mbf}[1]{\mathbf{#1}}
\newcommand{\tcr}[1]{\textcolor{black}{#1}}
\newcommand{\tcb}[1]{\textcolor{black}{#1}}
\newcommand{\etal}{\emph{et al.}}

\newcommand{\dctfac}{\bigg(\frac{\epsilon-1}{\epsilon}\bigg)}

\DeclareMathOperator\Tr{Tr}

\begin{document}

\title{Finite field formalism for bulk electrolyte solutions}

\author{Stephen J. Cox}
\affiliation{Department of Chemistry, University of Cambridge,
  Lensfield Road, Cambridge CB2 1EW, United Kingdom}
\email{sjc236@cam.ac.uk, ms284@cam.ac.uk}

\author{Michiel Sprik} 
\affiliation{Department of Chemistry, University of Cambridge,
  Lensfield Road, Cambridge CB2 1EW, United Kingdom}

\date{\today}

\begin{abstract}
  The manner in which electrolyte solutions respond to electric fields
  is crucial to understanding the behavior of these systems both at,
  and away from, equilibrium. The present formulation of linear
  response theory for such systems is inconsistent with common
  molecular dynamics (MD) implementations. Using the finite field
  formalism, suitably adapted for finite temperature MD, we
  investigate the response of bulk aqueous NaCl solutions to both
  finite Maxwell ($\mbf{E}$) and electric displacement ($\mbf{D}$)
  fields. The constant $\mbf{E}$ Hamiltonian allows us to derive the
  linear response relation for the ionic conductivity in a simple
  manner that is consistent with the forces used in conventional MD
  simulations. Simulations of a simple point charge model of an
  electrolyte solution at constant $\mbf{E}$ yield conductivities at
  infinite dilution within 15\,\% of experimental values. The finite
  field approach also allows us to measure the solvent's dielectric
  constant from its polarization response, which is seen to decrease
  with increasing ionic strength. Comparison of the dielectric
  constant measured from polarization response versus polarization
  fluctuations enables direct evaluation of the dynamic contribution
  to this dielectric decrement, which we find to be small but not
  insignificant. Using the constant $\mbf{D}$ formulation, we also
  rederive the Stillinger-Lovett conditions, which place strict
  constraints on the coupling between solvent and ionic polarization
  fluctuations.
\end{abstract}

\maketitle

\section{Introduction}
\label{sec:intro}

The role of ionic solutes in water is crucial across a broad range of
scientific and technological applications. One example is the
well-known Hofmeister series whereby simply changing the identity of
the ions has a profound effect on protein structure and stability (see
e.g. Ref.~\onlinecite{zhang2006interactions} for an overview). Along
similar lines, tuning the electrolyte is one route to controlling
self-assembly
\cite{min2008role,sukhorukov2001ph,kolny2002self,simard2000formation}
and the nucleation of molecular crystals
\cite{duff2014salting}. Recent work has also shown that ionic solutes
can significantly impact on the ice nucleating ability of
atmospherically relevant minerals \cite{whale2018enhancement}, which
is also likely to have consequences for controlling ice nucleation in
cryopreservation systems \cite{morris2013controlled}. Electrolyte
solutions are also important for energy storage applications
\cite{xia2017electrolytes}. This widespread importance of electrolytes
is a driving force for understanding their fundamental physical
behavior. In this regard, the microscopic insight offered by molecular
simulations makes them an increasingly important tool in addition to
experimental studies. However, the long-ranged nature of Coulombic
interactions poses major challenges for molecular simulations,
especially when used in conjunction with periodic boundary conditions
(PBC), which are typical for studies of condensed phase
systems. Originally developed for the study of ferroelectric
capacitors using Kohn-Sham density functional theory, the finite field
methods developed by Stengel \etal\cite{stengel2009electric} have
recently been extended to finite temperature molecular dynamics
simulations \cite{zhang2016computing1,zhang2016computing2}, and have
been shown to be an effective tool for dealing with the effects of
finite size when computing properties such as the capacitance of the
Helmholtz layer
\cite{zhang2016finite,sayer2017charge,sayer2019finite}.

The purpose of this article is to take the finite field methods
developed in
Refs.~\onlinecite{zhang2016computing1,zhang2016computing2,zhang2016finite}
for the study of dielectrics and solid/electrolyte interfaces and use
them to study bulk aqueous electrolyte solutions. This offers many
conceptual advantages over existing theoretical treatments. In
particular, we will find that the linear response (LR) relation for
the static conductivity can be derived in a much simpler form than in
existing formulations
\cite{caillol1986theoretical,caillol1989electricalI,caillol1989electricalII},
and in a manner consistent with common MD implementations. We will
also derive the Stillinger-Lovett (SL)
conditions\cite{stillinger1968ion,stillinger1968general} for bulk
electrolyte solutions in the presence of an imposed electric
displacement field. This approach offers a simplifying perspective
that readily lends itself to an intuitive physical understanding of
the observed anticorrelations between ionic and solvent polarization.

Much of our current understanding on this topic stems from the seminal
works of Caillol, Levesque and Weis
\cite{caillol1986theoretical,caillol1989electricalI,caillol1989electricalII}
(CLW). The foundation of their approach is the perturbative
Hamiltonian,
\begin{widetext}
\begin{equation}
  \label{eqn:HCLW}
  \Delta\mathcal{H}_{\rm CLW} =
  -c_{\rm em}^{-1}\int\!\mrm{d}\mbf{r}\,\mbf{j}_{\rm ion}(\mbf{r},t)\cdot\bm{\mathcal{A}}(\mbf{r},t)
  + \int\!\mrm{d}\mbf{r}\,\rho_{\rm ion}(\mbf{r},t)\varphi(\mbf{r},t)
  - \int\!\mrm{d}\mbf{r}\,\mbf{P}_{\rm wat}(\mbf{r},t)\cdot\bm{\mathcal{E}}(\mbf{r},t),
\end{equation}
\end{widetext}
where $\mbf{j}_{\rm ion}$ is the ionic current density, $\rho_{\rm
  ion}$ is the ionic charge density, $\mbf{P}_{\rm wat}$ is the
polarization of the solvent water molecules, and $c_{\rm em}$ is the
speed of light. The vector potential $\bm{\mathcal{A}}$ and the scalar
potential $\varphi$ are related to the `external' electric field by,
\begin{equation}
  \label{eqn:Epert-CLW}
  \bm{\mathcal{E}}(\mbf{r},t) = -c_{\rm em}^{-1}\partial_t \bm{\mathcal{A}}(\mbf{r},t)
  - \nabla\varphi(\mbf{r},t).
\end{equation}
The external field is the field that would be present in space
occupied by the sample if the sample were absent, see
e.g. Ref.~\onlinecite{MaddenKivelson1984sjc}. CLW work in a gauge in
which $\varphi = 0$. Note that this only applies to the perturbative
Hamiltonian $\Delta\mathcal{H}_{\rm CLW}$, and exploits the fact that
LR is gauge invariant \cite{caillol1986theoretical}. In what follows,
we draw heavily on the original work of CLW
\cite{caillol1986theoretical,caillol1989electricalI,caillol1989electricalII},
and we therefore save a detailed discussion for Sec.~\ref{sec:theory},
presenting instead just a brief summary of their approach here. First,
$\Delta\mathcal{H}_{\rm CLW}$ is used to express the external
susceptibilities $\{\bm{\chi}\}$ (see Eq.~\ref{eqn:chiPwP}) in terms
of the current and polarization fluctuations. Using Fulton's approach
\cite{fulton1978dipole}, $\bm{\mathcal{E}}$ is then expressed in terms
of the Maxwell field $\mbf{E}$, from which the ionic conductivity
$\sigma_{\rm ion}(\omega)$ is found from the constitutive relation,
Eq.~\ref{eqn:jion-constit}. The results of this procedure are
fluctuation (i.e. Green-Kubo) formulas appropriate for PBC with Ewald
summation, such as,
\begin{equation}
  \label{eqn:sigma_CLW}
  \sigma_{\rm ion}(0) =
  \frac{\beta}{3\Omega}\int_{0}^{\infty}\!\mrm{d}\tau\,\langle\mbf{J}_{\rm ion}(\tau)\cdot\mbf{J}_{\rm ion}(0)\rangle
\end{equation}
for the static ionic conductivity, where $\Omega$ is the volume of the
system, $\beta = 1/k_{\rm B}T$ ($k_{\rm B}$ is Boltzmann's constant,
and $T$ the temperature), and $\mbf{J}_{\rm ion} = \Omega\mbf{j}_{\rm
  ion}$ is the total ionic current. The ensemble average is understood
to be taken in the absence of the perturbing field.

The CLW formulation of LR is the basis of many simulation studies of
conducting liquids, and is used to derive the Einstein-Helfand
relation for the static conductivity \cite{schroder2008computation} as
well as to compute dielectric spectra
\cite{schroder2008collective,schroder2008computation,sega2013calculation,sega2014kinetic,rinne2014dissecting,sega2015kinetic}. However,
$\Delta\mathcal{H}_{\rm CLW}$ is not a Hamiltonian from which the
forces required for MD simulations can be readily derived. This is a
rather disconcerting aspect of this LR formulation, especially if we
wish to drive the simulated system with a perturbing field in a
rigorous manner. This is one of the issues we address in this
article. This also leads to practical benefits, as it allows us to
directly measure the solvent dielectric constant and ionic
conductivity directly from the response to a finite field, rather than
relying on Green-Kubo formulas that can be difficult to converge for
electrolyte systems (see Fig.~\tcr{S8}). Such an approach also allows
us to directly measure the `dynamic contribution' to the dielectric
decrement as the difference between the dielectric constant measured
from the solvent response, to that from its polarization fluctuations
at equilibrium. Since its conception by Hubbard and co-workers
\cite{hubbard1977kinetic,hubbard1977dielectric,hubbard1979molecular},
and later Felderhof \cite{felderhof1984dielectric}, the understanding
and calculation of this dynamic contribution has proved challenging
for both theory and simulation. Based on a linear hydrodynamics model,
Chandra \etal\cite{chandra1993dielectric} derived that this dynamic
decrement strictly vanishes for spherical ions in a solvent of
arbitrary molecular symmetry. Later simulations from Chandra
\cite{chandra2000static}, however, suggested that the dynamic
contribution was in fact non-zero, but still approximately two orders
of magnitude smaller than the equilibrium contributions to the
dielectric decrement. Results from our simulations corroborate this
later finding that the dynamic contribution is finite, although we
will see that it is significantly larger than suggested by Chandra.

In addition to providing a means to simulating systems at constant
Maxwell field $\mbf{E}$, the recent developments of
Refs.~\onlinecite{zhang2016computing1,zhang2016computing2,zhang2016finite}
also outline a procedure for performing simulations at fixed electric
displacement $\mbf{D}$. Methodologically, this is perhaps a more
significant theoretical advance than the constant $\mbf{E}$
ensemble. As such, the response of bulk electrolyte solutions to
constant $\mbf{D}$ fields are not widely studied with computer
simulation. However, we note that Caillol and co-workers realized that
$\epsilon^\prime=0$ is a relevant boundary condition in a formulation
in which the sample is surrounded by a medium of dielectric constant
$\epsilon^\prime$
\cite{caillol1989electricalI,caillol1989electricalII,caillol1994comments}. This
was later identified as corresponding to a $\mbf{D}=\mbf{0}$ ensemble
\cite{zhang2016computing1,zhang2016computing2,zhang2016finite}, which
is elaborated upon in Sec.~\ref{subsec:SL-conditions}. One example of
previous work that has explicitly used the constant $\mbf{D}$
formulation for electrolyte solutions is that of Pache and Schmid
\cite{pache2018molecular} who investigated the concentration
dependence of the solvent dielectric constant of various electrolyte
solutions, and reported severe dielectric saturation at moderate field
strengths (e.g. the dielectric constant decreases by approx. 50\% for
$\mbf{D}=2$\,V/\AA). These results are discussed in the context of the
SL conditions below.

The rest of the article is as follows. In Sec.~\ref{sec:theory} we
give a general outline of the relevant theory, with derivations of the
LR relation for the static conductivity given in
Sec.~\ref{subsec:LR-sigma}, and the SL conditions in
Sec.~\ref{subsec:SL-conditions}. Simulation methods are given in
Sec.~\ref{sec:methods}. In Sec.~\ref{sec:sim} we present results from
our simulations, with the response to finite $\mbf{E}$ and finite
$\mbf{D}$ given in Secs.~\ref{subsec:sim-E} and~\ref{subsec:sim-D},
respectively. We end with a summary in Sec.~\ref{sec:summary}.

\section{Theoretical outline}
\label{sec:theory}

Central to this work are the constitutive relations that relate the
Maxwell field $\mbf{E}$ to the water polarization $\mbf{P}_{\rm wat}$
and the ionic current density $\mbf{j}_{\rm ion}$. They are,
\begin{align}
  \mbf{P}_{\rm wat}(\mbf{r},\omega) &= \frac{\epsilon_{\rm wat}(\omega)-1}{4\pi}\mbf{E}(\mbf{r},\omega), \label{eqn:Pwat-constit} \\[7pt]
  \mbf{j}_{\rm ion}(\mbf{r},\omega) &= \sigma_{\rm ion}(\omega)\mbf{E}(\mbf{r},\omega), \label{eqn:jion-constit}
\end{align}
where $\epsilon_{\rm wat}$ is the dielectric constant of the solvent
water, and $\sigma_{\rm ion}$ is the ionic conductivity. $\mbf{P}_{\rm
  wat}$ and $\mbf{j}_{\rm ion}$ are understood to be ensemble
averages. For notational convenience, we omit angled brackets when
this is clear from context, although the standard notation
`$\langle\cdot\rangle$' will be used to denote such ensemble averages
when required. In writing Eqs.~\ref{eqn:Pwat-constit}
and~\ref{eqn:jion-constit}, the possibility of time dependent fields
has been considered, with $\omega$ the frequency of
oscillation. Throughout the article, the Fourier transform in the time
domain is defined as e.g.,
\begin{equation}
  \label{eqn:FLT}
  \mbf{P}_{\rm wat}(\mbf{r},\omega) = \int_{-\infty}^{\infty}\!\mrm{d}t\,\mbf{P}_{\rm wat}(\mbf{r},t)\exp\left(i\omega t\right).
\end{equation}
(Note that this is a full transform over the time domain. For the
susceptibilities this would be a half-transform.) From a molecular
simulation, one has direct access to $\mbf{P}_{\rm wat}$ and
$\mbf{j}_{\rm ion}$. Experimentally, however, it is the total current
$\mbf{j} = \mbf{j}_{\rm ion} + \partial_t \mbf{P}_{\rm wat}$ that is
measured. Its relation to $\mbf{E}$ is,
\begin{equation}
  \mbf{j}(\omega) = \sigma_{\rm T}(\omega)\mbf{E}(\omega), \label{eqn:jtot-constit}
\end{equation}
with,
\begin{equation}
  \label{eqn:sigmaT}
  \sigma_{\rm T}(\omega) = \sigma_{\rm ion}(\omega) - \frac{i\omega}{4\pi}\big[\epsilon_{\rm wat}(\omega)-1\big].
\end{equation}
Note that in writing Eqs.~\ref{eqn:Pwat-constit},
\ref{eqn:jion-constit} and~\ref{eqn:jtot-constit}, we have implicitly
assumed that the response is local.

In addition to these constitutive relations, it is also desirable to
relate $\mbf{P}_{\rm wat}$, $\mbf{j}_{\rm ion}$ and $\mbf{j}$ to the
perturbing field. For example,
\begin{align}
  4\pi\mbf{P}_{\rm wat}(\omega) &= \bm{\chi}_{P_{\rm w}P}(\omega)\ast\mbf{E}_0(\omega), \label{eqn:chiPwP}
\end{align}
Similar relations are also defined for $\mbf{j}_{\rm ion}$, $\mbf{j}$
and the total polarization $\mbf{P} = \mbf{P}_{\rm ion} + \mbf{P}_{\rm
  wat}$, each with its own susceptibility ($\bm{\chi}_{j_{\rm i}P}$,
$\bm{\chi}_{jP}$ and $\bm{\chi}_{PP}$, respectively), which will be
referred to collectively as $\{\bm{\chi}\}$.\footnote{With the Fourier
  transform defined by Eq.~\ref{eqn:FLT}, the complex susceptibility
  can be expressed as $\bm{\chi}(\omega) = \bm{\chi}^\prime(\omega) +
  i\bm{\chi}^{\prime\prime}(\omega)$, where $\bm{\chi}^\prime$ and
  $\bm{\chi}^{\prime\prime}$ denote the real and imaginary parts of
  $\bm{\chi}$, respectively. This follows,
  e.g. Refs.~\onlinecite{MaddenKivelson1984sjc,HANSEN2013265}. Another
  convention often seen in the literature is to define the Fourier
  transform with a minus sign in the exponent. In this case,
  $\bm{\chi}(\omega) = \bm{\chi}^\prime(\omega) -
  i\bm{\chi}^{\prime\prime}(\omega)$.} Eq.~\ref{eqn:chiPwP} also
introduces the shorthand notation in which `$\ast$' denotes both
tensor contraction and spatial convolution, i.e.,
\begin{equation}
  \bm{\chi}(\omega)\ast\mbf{E}_0(\omega) =
  \sum_{\alpha\gamma}\int\!\mrm{d}\mbf{r}^\prime\,
  \chi_{\alpha\gamma}(|\mbf{r}-\mbf{r}^{\prime}|,\omega)E_{0,\gamma}(\mbf{r}^{\prime},\omega)\hat{\mbf{e}}_\alpha,
\end{equation}
where $\alpha$ and $\gamma$ denote components of a Cartesian
coordinate system, and $\hat{\mbf{e}}_\alpha$ is the unit vector along
direction $\alpha$. Our choice of notation `$\mbf{E}_0$' for the
perturbing field requires some clarification. For consistency with the
finite field Hamiltonians (see Eqs.~\ref{eqn:H-E} and~\ref{eqn:H-D}),
we ultimately wish to identify $\mbf{E}_0$ with either the Maxwell
field $\mbf{E}$ or displacement field $\mbf{D}$. For the constant
$\mbf{E}$ ensemble, this is straightforward. Imposing constant
$\mbf{D}$, on the other hand, gives rise to subtleties which motivates
the use of the following general Hamiltonian to formulate the LR
relations,
\begin{align}
  \mathcal{H}(\mbf{r}^{N},\mbf{p}^N) &= \mathcal{H}_0(\mbf{p}^{N},\mbf{r}^N) - \int\!\mrm{d}\mbf{r}\,\mbf{E}_0\cdot\mbf{P}(\mbf{r}^{N}), \\[7pt]
  &= \mathcal{H}_0(\mbf{r}^{N},\mbf{p}^N) - \Omega\mbf{E}_0\cdot\mbf{P}(\mbf{r}^{N}). \label{eqn:genH-LR}
\end{align}
In going from the first to the second lines, we note that we only
consider uniform fields. $\mathcal{H}_0$ is the Hamiltonian when
$\mbf{E}_{0}=\mbf{0}$. Both $\mbf{E}_0$ and $\mathcal{H}_0$ depend
upon the choice of boundary conditions.

\subsection{Linear response relation for the static conductivity}
\label{subsec:LR-sigma}

As discussed in Sec.~\ref{sec:intro}, the CLW approach to LR is based
on the perturbative Hamiltonian given by Eq.~\ref{eqn:HCLW}. This
couples $\mbf{j}_{\rm ion}$ to the vector potential
$\bm{\mathcal{A}}$. In comparison, within the finite field
formulation, the Hamiltonian for an imposed, uniform, although
potentially time-dependent, $\mbf{E}$ reads,
\begin{equation}
    \mathcal{H}_{\mbf{E}}(\mbf{r}^N,\mbf{p}^N) = \mathcal{H}_{\rm PBC}(\mbf{r}^N,\mbf{p}^N) - \frac{\Omega}{8\pi}|\mbf{E}|^2 - \Omega\mbf{E}\cdot\mbf{P}(\mbf{r}^N), \label{eqn:H-E}
\end{equation}
where $\mathcal{H}_{\rm PBC}$ is a `standard' Hamiltonian used in MD
simulation, which comprises all interatomic interactions, with
electrostatic interactions calculated with Ewald summation (or one of
its mesh based variants). Again we emphasize that $\mbf{E}$ is the
Maxwell field. The total polarization is defined as the time integral
of the current,
\begin{equation}
  \label{eqn:P-itinerant}
  \mbf{P}(\mbf{r}^N) = \frac{1}{\Omega}\sum_{i}q_i\mbf{r}_{i}(t),
\end{equation}
where $q_i$ is the charge of atom $i$, whose position at time $t$ is
denoted by $\mbf{r}_{i}(t)$. Crucially, the sum includes both the
atoms of the solvent molecules and the charged ions. The polarization
that couples to $\mbf{E}$ is the itinerant polarization. When an ion
leaves the primary simulation cell, we follow its position out of the
box, and it is these coordinates that enter into the sum in
Eq.~\ref{eqn:P-itinerant}. Use of the itinerant polarization in
molecular simulations with PBC is not new; as noted by Caillol, it is
the itinerant polarization that enters naturally in the Ewald sum, and
satisfies key statistical mechanical properties such as the SL sum
rules \cite{caillol1994comments}. Although Eq.~\ref{eqn:H-E} was first
derived on thermodynamic grounds, we stress that it is a full
microscopic Hamiltonian. This was recently formalized in
Ref.~\onlinecite{sprik2018finite} where is was derived from an
extended Lagrangian based on arguments of theoretical mechanics. In
fact, the finite field Hamiltonians given by Eqs.~\ref{eqn:H-E}
and~\ref{eqn:H-D} can be obtained by a Power-Zienau gauge
transformation \cite{power1959coulomb} from the minimal coupling
Hamiltonian used by CLW, with the restriction that $\mbf{E}$ and
$\mbf{D}$ must be uniform. These Hamiltonians also respect the
inherently multivalued nature of the polarization under PBC
\cite{zhang2016finite,sprik2018finite}.

With the form of $\mathcal{H}_{\mbf{E}}$ given by Eq.~\ref{eqn:H-E},
we can readily identify the perturbing field $\mbf{E}_0$ with the
Maxwell field $\mbf{E}$ (see Eq.~\ref{eqn:genH-LR}). Note that the
second term on the right hand side of Eq.~\ref{eqn:H-E} is constant
for a given $\mbf{E}$; it is required to ensure that the electric
enthalpy at constant $\mbf{E}$ and electric internal energy at
constant $\mbf{D}$ are each other's Legendre transforms
\cite{landau&lifshitz_EDCM,stengel2009electric,zhang2016computing1,sprik2018finite}.
The derivation of the LR relation for $\sigma_{\rm ion}(0)$ now
follows standard textbook arguments \cite{HANSEN2013265}. Taking
$\mbf{E}$ to be a monochromatic field of frequency $\omega$, aligned
along the $x$ direction for convenience, the total current is,
\begin{equation}
  \langle J_x(t)\rangle = \chi_{JM}(\omega)E\exp(-i\omega t),
\end{equation}
with,
\begin{equation}
  \chi_{JM}(\omega) = \beta\int_0^{\infty}\!\mrm{d}\tau\,\langle J_x(\tau)J_{x}(0)\rangle\exp(i\omega \tau).
\end{equation}
Comparing to Eq.~\ref{eqn:jtot-constit} we find,
\begin{equation}
  \sigma_{\rm T}(\omega) = \frac{\beta}{\Omega}\int_0^{\infty}\!\mrm{d}\tau\,\langle J_x(\tau)J_{x}(0)\rangle\exp(i\omega \tau).
\end{equation}
Exploiting the isotropy of the system, and taking the static limit
(see Eq.~\ref{eqn:sigmaT}) gives,
\begin{equation}
  \lim_{\omega\to 0} \sigma_{\rm T}(\omega) = \sigma_{\rm ion}(0) = \frac{\beta}{3\Omega}\int_0^{\infty}\!\mrm{d}\tau\,\langle \mbf{J}(\tau)\cdot\mbf{J}(0)\rangle. \label{eqn:sigma_SSV}
\end{equation}

Aside from being a decidedly simpler derivation than that based on
$\Delta\mathcal{H}_{\rm CLW}$
\cite{caillol1986theoretical,caillol1989electricalI,caillol1989electricalII},
the pleasing aspect of the above derivation is that
$\mathcal{H}_{\mbf{E}}$ is the same Hamiltonian used to derive the
forces for MD simulations. It amounts to simply adding a force
$\mbf{f}_{\mbf{E}} = q_i\mbf{E}$ to each atom $i$ in the
simulation. We exploit this fact in our simulations, which are
presented in Sec.~\ref{subsec:sim-E}. Note that the use of uniform
fields is crucial to this formulation.

\subsection{Stillinger-Lovett conditions}
\label{subsec:SL-conditions}

The Stillinger-Lovett conditions are a statement that the mobile ions
completely screen the solvent from slowly-varying, static fields
\cite{stillinger1968ion,stillinger1968general,blum1982perfect,martin1988sum,carnie1983sum,caillol1989electricalI,caillol1989electricalII,caillol1994comments}. From
the finite field Hamiltonian for constant displacement field
$\mbf{D}$,
\begin{equation}
  \mathcal{H}_{\mbf{D}}(\mbf{r}^N,\mbf{p}^N) =
  \mathcal{H}_{\rm PBC}(\mbf{r}^N,\mbf{p}^N) + \frac{\Omega}{8\pi}|\mbf{D}-4\pi\mbf{P}(\mbf{r}^N)|^2 \label{eqn:H-D},
\end{equation}
it is possible to formulate the SL conditions in a manner that
ultimately avoids invoking abstract cavity relations. In what follows,
we will work in an ensemble in which the displacement field is fixed
in all three Cartesian directions, $\mbf{D} = D_x\hat{\mbf{x}} +
D_y\hat{\mbf{y}} + D_z\hat{\mbf{z}}$. We begin by stating the LR
relation for the total polarization in response to $\mbf{D}$,
\begin{equation}
  \langle \mbf{P}\rangle_{\mbf{D}} = \frac{\beta\Omega}{3}\langle|\delta \mbf{P}|^2\rangle_0 \mbf{D}, \label{eqn:LR-constD}
\end{equation}
where $\delta \mbf{P} = \mbf{P} - \langle \mbf{P}\rangle_0$. The
subscript `0' indicates averages taken at $\mbf{D}=\mbf{0}$. From the
definition of the polarizability, the fluctuations in the total dipole
moment $\mbf{M} = \Omega \mbf{P}$ are related to the dielectric
constant \cite{zhang2016computing2},
\begin{equation}
  \frac{4\pi\beta}{3\Omega}\langle|\delta \mbf{M}|^2\rangle_0 = \dctfac. \label{eqn:SL-dielectric}
\end{equation}
For conducting electrolyte systems, we are interested in the limit
$\epsilon\to\infty$,
\begin{equation}
  \frac{4\pi\beta}{3\Omega}\langle|\delta \mbf{M}|^2\rangle_0 = 1. \label{eqn:SL-full1}
\end{equation}
Substituting Eq.~\ref{eqn:SL-full1} into Eq.~\ref{eqn:LR-constD} gives
$4\pi\langle \mbf{P}\rangle_{\mbf{D}} = \mbf{D}$. From the fundamental
equation of Maxwell's theory of dielectrics, $\mbf{D} = \mbf{E} +
4\pi\mbf{P}$, we find,
\begin{equation}
  \langle \mbf{E}\rangle = 0.
\end{equation}
This simply reflects the fact that at equilibrium, the total electric
field inside a conducting medium vanishes. From
Eqs.~\ref{eqn:Pwat-constit} and~\ref{eqn:jion-constit} we find that,
$\langle \mbf{P}_{\rm wat}\rangle_{\mbf{D}} = \langle \mbf{j}_{\rm
  ion}\rangle_{\mbf{D}} = \mbf{0}$.

Following the discussion at the end of Sec.~\ref{sec:theory}, we now
consider the general Hamiltonian given by Eq.~\ref{eqn:genH-LR}. In
this case, the external susceptibilities $\{\bm{\chi}\}$ are related
to the time correlation functions of the system at
$\mbf{E}_0=\mbf{0}$,
\begin{align}
  \bm{\chi}_{P_{\rm w}P}(\omega) &= 4\pi\beta\langle\mbf{P}_{\rm wat}(t)\mbf{j}(0)\rangle_{\omega}, \label{eqn:chiPwP_tcf}  \\[7pt]
  \bm{\chi}_{j_{\rm i}P}(\omega) &= 4\pi\beta\langle\mbf{j}_{\rm ion}(t)\mbf{j}(0)\rangle_{\omega}, \label{eqn:chijiP_tcf} \\[7pt]
  \bm{\chi}_{jP}(\omega)      &=  4\pi\beta\langle\mbf{j}(t)\mbf{j}(0)\rangle_{\omega}, \label{eqn:chijP_tcf}
\end{align}
where $\langle\cdot\rangle_{\omega}$ denotes the Fourier-Laplace
transform. (In writing Eqs.~\ref{eqn:chiPwP_tcf}--\ref{eqn:chijP_tcf},
we have used the fact that the average polarization and current
densities formally vanish at zero field.) Noting that $\mbf{j} =
\mbf{j}_{\rm ion} + \mbf{j}_{\rm wat}$, exploiting well-known
properties of time correlations functions, and integrating by parts a
number of times, we find,
\begin{widetext}
\begin{align}
  \bm{\chi}_{P_{\rm w}P}(\omega) &=
  4\pi\beta\bigg[\langle\mbf{P}_{\rm wat}\mbf{P}_{\rm wat}\rangle + \langle\mbf{P}_{\rm wat}\mbf{P}_{\rm ion}\rangle
    + i\omega\langle\mbf{P}_{\rm wat}(t)\mbf{P}_{\rm ion}(0)\rangle_{\omega}+ i\omega\langle\mbf{P}_{\rm wat}(t)\mbf{P}_{\rm wat}(0)\rangle_{\omega}\bigg], \label{eqn:chiPwP_pol}  \\[7pt]
  \bm{\chi}_{j_{\rm i}P}(\omega) &=
  -4\pi\beta i\omega\bigg[\langle\mbf{P}_{\rm ion}\mbf{P}_{\rm ion}\rangle + \langle\mbf{P}_{\rm ion}\mbf{P}_{\rm wat}\rangle
   + i\omega\langle\mbf{P}_{\rm ion}(t)\mbf{P}_{\rm wat}(0)\rangle_{\omega} + i\omega\langle\mbf{P}_{\rm ion}(t)\mbf{P}_{\rm ion}(0)\rangle_{\omega}\bigg], \label{eqn:chijiP_pol} \\[7pt]
  \bm{\chi}_{jP}(\omega)      &=  \bm{\chi}_{j_{\rm i}P}(\omega) - i\omega\bm{\chi}_{P_{\rm w}P}(\omega). \label{eqn:chijP_pol}
\end{align}
\end{widetext}
As the time derivative of the polarization gives the current,
$\mbf{j}(\omega) = -i\omega\mbf{P}(\omega)$,
\begin{equation}
  \bm{\chi}_{PP}(\omega) = \frac{i}{\omega}\bm{\chi}_{jP}(\omega) =
  \bigg[\bm{\chi}_{P_{\rm w}P}(\omega) + \frac{i}{\omega}\bm{\chi}_{j_{\rm i}P}(\omega)\bigg]. \label{eqn:chiPP_pol0}
\end{equation}
Importantly, for static fields,
\begin{equation}
  \bm{\chi}_{PP}(\omega=0) = \bigg[\bm{\chi}_{P_{\rm w}P}(\omega=0) + \lim_{\omega\to 0}\frac{i}{\omega}\bm{\chi}_{j_{\rm i}P}(\omega)\bigg]. \label{eqn:chiPP_static}
\end{equation}
As both $\langle \mbf{P}_{\rm wat}\rangle_{\mbf{D}}$ and $\langle
\mbf{j}_{\rm ion}\rangle_{\mbf{D}}$ vanish, consistency demands that,
\begin{align}
  \int_{\Omega}\!\mrm{d}\mbf{r}^{\prime}\,\bm{\chi}_{P_{\rm w}P}(\mbf{r},\mbf{r}^\prime,\omega=0) = \mbf{0} \label{eqn:chiPwP_int0} \\[7pt]
  \int_{\Omega}\!\mrm{d}\mbf{r}^{\prime}\,\bm{\chi}_{j_{\rm i}P}(\mbf{r},\mbf{r}^\prime,\omega=0) = \mbf{0} \label{eqn:chijiP_int0}
\end{align}
From Eqs.~\ref{eqn:chiPwP_pol} and~\ref{eqn:chiPwP_int0} we derive the
first SL sum rule,
\begin{equation}
  \frac{4\pi\beta}{3\Omega}\bigg[\langle|\mbf{M}_{\rm wat}|^2\rangle + \langle\mbf{M}_{\rm wat}\cdot\mbf{M}_{\rm ion}\rangle\bigg] = 0. \label{eqn:SL-wat}
\end{equation}

According to Eq.~\ref{eqn:chijiP_pol}, $\bm{\chi}_{j_{\rm
    i}P}(\omega=0) = 0$, and Eq.~\ref{eqn:chijiP_int0} therefore
contains no useful information. In order to proceed, we will follow
Fulton's approach \cite{fulton1978dipole} to relate $\mbf{E}$ to
$\mbf{E}_0$ and $\mbf{P}$,
\begin{equation}
  \mbf{E} = \mbf{E}_0 + 4\pi\mbf{G}_0\ast\mbf{P}, \label{eqn:FultonE}
\end{equation}
where $\mbf{G}_0$ is the Green's function for the constant $\mbf{D}$
ensemble. That is,
$\mbf{G}_0(\mbf{r},\mbf{r}^\prime)\ast{\bm\mu}^\prime$ is the electric
field at $\mbf{r}$ caused by a dipole ${\bm\mu}^\prime$ at position
$\mbf{r}^\prime$ under PBC at constant $\mbf{D}$. As noted in
Sec.~\ref{sec:intro}, it has already been established that the
$\mbf{D}=\mbf{0}$ Hamiltonian has the same structure as that derived
from a reaction field approach in which the surrounding medium has
vanishing dielectric constant ($\epsilon^\prime=0$). We can exploit
this fact to obtain $\mbf{G}_0$, which is readily achieved by
following Ref.~\onlinecite{caillol1992asymptotic}. For the isotropic
systems considered here, we only require its trace at $\omega=0$,
\begin{equation}
  \label{eqn:TrG0}
  \Tr{\mbf{G}_0(\mbf{r},\mbf{r}^\prime,\omega=0)} = -\bigg[\delta_{\rm EW}(\mbf{r}-\mbf{r}^\prime) + \frac{2}{\Omega}\bigg],
\end{equation}
where $\delta_{\rm EW}(\mbf{r}-\mbf{r}^\prime)$ is the periodic Dirac
delta-function. We now substitute $4\pi\mbf{P} =
\bm{\chi}_{PP}\ast\mbf{E}_0$ into Eq.~\ref{eqn:FultonE}, which after
setting $\mbf{E}=\mbf{0}$ gives,
\begin{equation}
  \label{eqn:fulton-rearr}
  \int_{\Omega}\!\mrm{d}\mbf{r}^\prime\int_{\Omega}\!\mrm{d}\mbf{r}^{\prime\prime}\,\mbf{G}_0(\mbf{r},\mbf{r}^\prime)\cdot\bm{\chi}_{PP}(\mbf{r}^\prime,\mbf{r}^{\prime\prime},\omega=0) = -\mbf{1}.
\end{equation}
Equations~\ref{eqn:chiPP_static} and~\ref{eqn:chiPwP_int0} allow us to
write the left hand side as,
\begin{equation}
  \int_{\Omega}\!\mrm{d}\mbf{r}^\prime\int_{\Omega}\!\mrm{d}\mbf{r}^{\prime\prime}\,\mbf{G}_0(\mbf{r},\mbf{r}^\prime)\cdot\lim_{\omega\to 0}\frac{i}{\omega}\bm{\chi}_{j_{\rm i}P}(\mbf{r}^\prime,\mbf{r}^{\prime\prime},\omega).
\end{equation}
The fact that the system is isotropic allows us to write
$\bm{\chi}_{j_{\rm i}P} = \chi_{j_{\rm i}P}\mbf{1}$, where
$\chi_{j_{\rm i}P}$ is a scalar. Moreover, as we are only concerned
with uniform fields, only the zero mode of the susceptibility
contributes upon integration. Thus, upon taking the trace we obtain,
%
\begin{widetext}
\begin{align}
  -\lim_{\omega\to 0}\int_{\Omega}\!\mrm{d}\mbf{r}^\prime\int_{\Omega}\!\mrm{d}\mbf{r}^{\prime\prime}\,\bigg[\delta_{\rm EW}(\mbf{r}-\mbf{r}^\prime) + \frac{2}{\Omega}\bigg]\frac{i}{\omega}\chi_{j_{\rm i}P}(\mbf{r}^\prime,\mbf{r}^{\prime\prime},\omega)
  &= -\lim_{\omega\to 0}\frac{3}{\Omega}\int_{\Omega}\!\mrm{d}\mbf{r}^\prime\int_{\Omega}\!\mrm{d}\mbf{r}^{\prime\prime}\,\frac{i}{\omega}\chi_{j_{\rm i}P}(\mbf{r}^\prime,\mbf{r}^{\prime\prime},\omega)
\end{align}
\end{widetext}
Using our expression for $\bm{\chi}_{j_{\rm i}P}$ in terms of the time
correlation functions (Eq.~\ref{eqn:chijiP_pol}), and after taking the
trace of the unit tensor (see Eq.~\ref{eqn:fulton-rearr}), we find,
\begin{equation}
  \frac{4\pi\beta}{\Omega}\int_{\Omega}\!\mrm{d}\mbf{r}^\prime\int_{\Omega}\!\mrm{d}\mbf{r}^{\prime\prime}\,\bigg[\langle\mbf{P}_{\rm ion}\cdot\mbf{P}_{\rm ion}\rangle + \langle\mbf{P}_{\rm ion}\cdot\mbf{P}_{\rm wat}\rangle\bigg] = 3,
\end{equation}
or,
\begin{align}
  \label{eqn:SL-ion}
  \frac{4\pi\beta}{3\Omega}\bigg[\langle|\mbf{M}_{\rm ion}|^2\rangle + \langle\mbf{M}_{\rm ion}\cdot\mbf{M}_{\rm wat}\rangle\bigg] = 1.
\end{align}
This is the second SL condition. As a sanity check, addition of
Eqs.~\ref{eqn:SL-wat} and~\ref{eqn:SL-ion} recovers the LR condition
for the total polarization fluctuations, Eq.~\ref{eqn:SL-full1}.

The above derivation is broadly similar to that of Caillol, Levesque
and Weis \cite{caillol1989electricalI}. The main difference lies in
the fact that we have associated $\mbf{G}_0$ as the Greens' function
for the constant-$\mbf{D}$ ensemble. The significance of this
statement becomes clear when we attempt to perturb the system with
finite $\mbf{E}_0$, which we show below that we can identify with
$\mbf{D}$. In the original CLW formulation, the perturbing field is
$\bm{\mathcal{E}}$ (Eq.~\ref{eqn:HCLW}), which is then associated with
the `cavity field' \cite{caillol1989electricalI}. The drawback of this
approach is that one cannot readily identify the molecular forces
associated with the perturbing field. In contrast, the current
formulation allows us to readily derive the appropriate forces from
$\mathcal{H}_{\mbf{D}}$ (Eq.~\ref{eqn:H-D}), and thus probe the
system's response to finite values of $\mbf{D}$. Crucial to the
current formulation is that we have accounted for the ions' charge in
the intinerant polarization, and the only source of $\mbf{D}$ is the
charge on the electrodes at infinity. The extent to which the SL
conditions remain valid at finite $\mbf{D}$ places this interpretation
on a firm statistical mechanical basis that can be verified
empirically by simulation, beyond the theoretical arguments in
Ref.~\onlinecite{sprik2018finite}.

In passing, we note that in a later paper \cite{caillol1994comments},
Caillol suggests that the itinerant polarization behaves like an
independent harmonic oscillator (Eq.~\ref{eqn:SL-full1}), and in that
sense, the total SL condition is a mere consequence of energy
equipartition. Such an interpretation is misleading when considered
more generally. The constant $\mbf{D}$ Hamiltonian is oblivious to
properties of the system; it simply does not know if it governs the
dynamics of a conductor or a dielectric. Arguments based on energy
equipartition would suggest Eq.~\ref{eqn:SL-full1} holds in all
cases. This, however, would violate the LR relations for dielectrics
(see Eq.~\ref{eqn:SL-dielectric}), which have previously been shown to
work well for describing bulk water \cite{zhang2016computing1}.

What remains is to establish that we can indeed associate the
perturbing field $\mbf{E}_{0}$ with the displacement field
$\mbf{D}$. To this end, we expand the quadratic term in
$\mathcal{H}_{\mbf{D}}$ (Eq.~\ref{eqn:H-D}),
\begin{equation}
  \mathcal{H}_{\mbf{D}} = \mathcal{H}_{\rm PBC} + 2\pi\Omega|\mbf{P}|^2
  + \frac{\Omega}{8\pi}|\mbf{D}|^2 - \Omega\mbf{D}\cdot\mbf{P}.
\end{equation}
Comparing to the general LR Hamiltonian (Eq.~\ref{eqn:genH-LR}), we
see that unlike the constant-$\mbf{E}$ ensemble, $\mathcal{H}_0$ (the
Hamiltonian at $\mbf{D}=\mbf{0}$) contains a term quadratic in the
total polarization. This is, of course, the origin of the distinctly
different fluctuations in the two ensembles
\cite{zhang2016computing1}. Once we acknowledge the belonging of the
$|\mbf{P}|^2$ term to $\mathcal{H}_0$, direct comparison of
Eqs.~\ref{eqn:genH-LR} and~\ref{eqn:H-D} allows us to identify
$\mbf{D}$ with $\mbf{E}_0$. 

\begin{figure*}[t]
  \includegraphics[width=12.08cm]{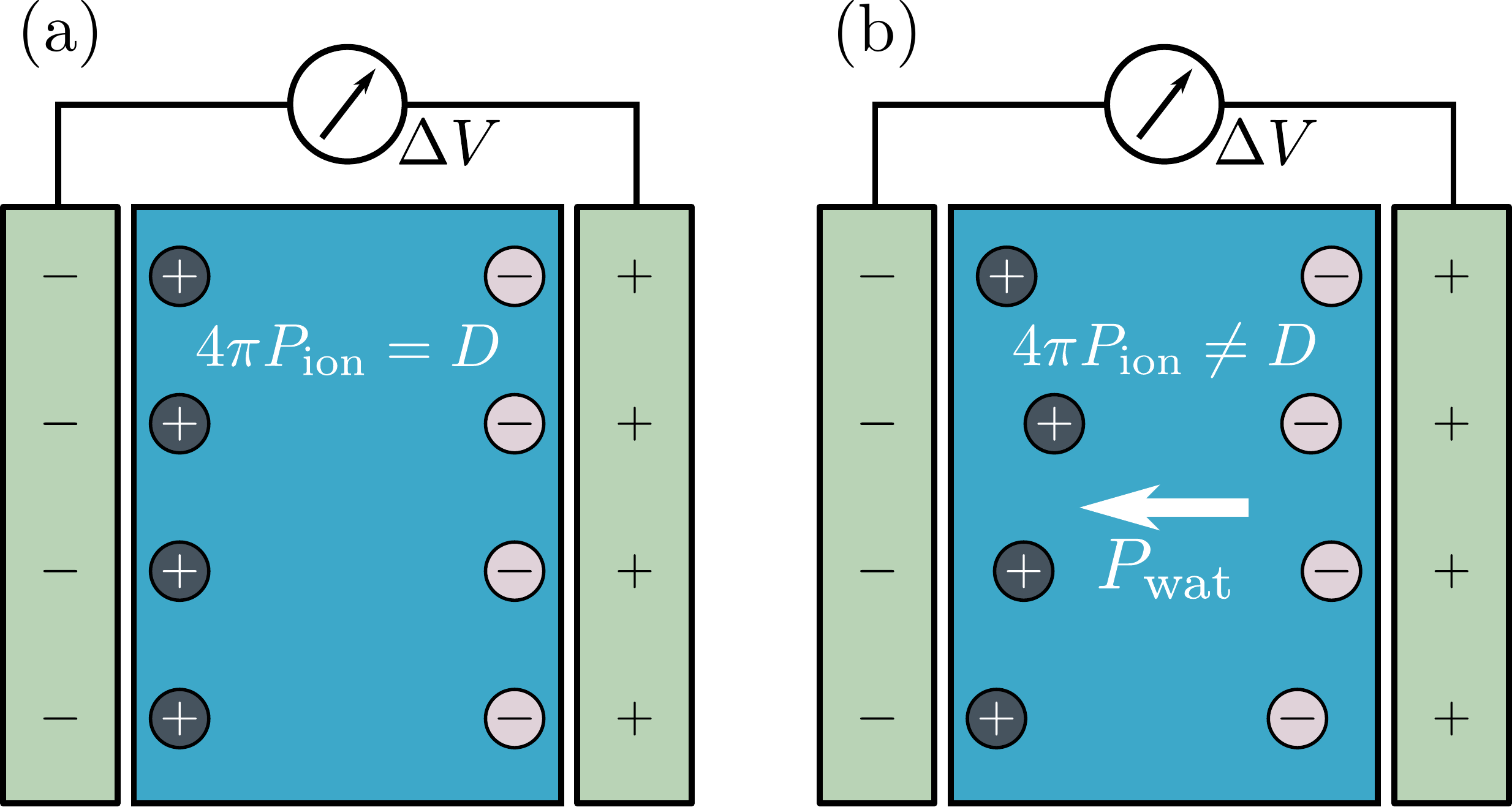}
  \caption{Schematic of electrolyte behavior at constant $\mbf{D} =
    D_x\hat{\mbf{x}}$, which can be modeled as a pair of electrodes
    held at constant charge. (a) At equilibrium, the ions relax such
    that $4\pi\langle\mbf{P}_{\rm ion}\rangle = \mbf{D}$. This total
    screening of the $\mbf{D}$ field by the ions means the average
    solvent polarization vanishes. (b) A thermal fluctuation displaces
    the ions from their equilibrium configuration, and the field is no
    longer completely screened. A transient solvent polarization is
    observed, which is in the opposite direction to the ionic
    polarization fluctuation.}
  \label{fig:SL-schematic}
\end{figure*}

We end this section with a comment on the coupling between ionic and
solvent polarization fluctuations. The SL conditions place strict
requirements on the behavior of the electrolyte solution. In
particular, it is clear that $\langle\delta\mbf{M}_{\rm
  wat}\cdot\delta\mbf{M}_{\rm ion}\rangle < 0$ if Eq.~\ref{eqn:SL-wat}
is to be satisfied. In other words, the ionic and water polarization
fluctuations are anticorrelated. The finite field method framework
provides a useful physical interpretation for this result. To this
end, we consider `hybrid' boundary conditions in which the
displacement field is only set along one of the Cartesian directions,
e.g. $\mbf{D} = D_x\hat{\mbf{x}}$, while tin-foil boundary conditions
are used in the transverse directions ($E_y = E_z = 0$). In this
constant-$D_x$ ensemble, the behavior is analogous to that of a system
between a pair of electrodes whose equal-and-opposite charges are
fixed \cite{zhang2016computing1}. At equilibrium, the ions relax such
that $4\pi\langle P_{x,\rm ion}\rangle = D_x$, and $\langle P_{x, \rm
  wat}\rangle = 0$. This is depicted schematically in
Fig.~\ref{fig:SL-schematic}\,(a). Owing to thermal motion, the ions
will fluctuate around their equilibrium configurations such that at
any instant $4\pi P_{x, \rm ion} \neq D_x$, and the field is not fully
screened, as shown in Fig.~\ref{fig:SL-schematic}\,(b). It is
reasonable to assume that the timescale on which the solvent
reorganizes is faster than that for the ions to relax back toward
equilibrium. Consequently, it is expected that a transient
polarization of the water will be observed, and aligned in the
opposite direction to the transient fluctuation in the ionic
polarization. While this parallel plate capacitor analogy cannot be
rigorously extended to the ensemble in which $\mbf{D} =
D_x\hat{\mbf{x}} + D_y\hat{\mbf{y}} + D_z\hat{\mbf{z}}$
\cite{zhang2016computing1}, we show in the S.I. that empirically the
SL conditions are still satisfied in the case of hybrid boundary
conditions. Thus, the underlying principle---that fluctuations in the
ionic polarization lead to a transient incomplete screening with
associated solvent response---appears to hold true in both ensembles.

\section{Methods}
\label{sec:methods}

The system we consider is aqueous sodium chloride (\ce{NaCl}) with
concentrations in the range $0.05 \lesssim c \lesssim 8.65$\,M,
modeled under 3D PBC with Ewald summation, and in the absence of
extended interfaces. For $c>0.4$\,M each simulation comprised 256
SPC/E \cite{BerendsenStraatsma1987sjc} water molecules, corresponding
to a number of ion pairs between two and forty for the concentration
ranges investigated. The cubic dimension of the simulation cell was
$L=19.73$\,\AA, and the volume $\Omega = L^3$ was the same for all
$c>0.4$\,M. This corresponds to a constant number density of water of
$33.33\times10^{-3}$\,\AA$^{-3}$. Consequently, the pressure increases
dramatically as the concentration of the solution increases. However,
we show in the S.I. that similar behavior is observed when we adjust
the size of the simulation cell such that the average pressure is
approximately constant. For the two lowest concentrations
investigated, $c\approx 0.05$\,M and $c\approx 0.11$\,M, simulations
were performed with 1024 and 2048 water molecules, respectively, with
two ion pairs in both cases. The cell lengths were $L=31.32$\,\AA{}
(0.05\,M) and $L=39.46$\,\AA{} (0.11\,M). \tcb{As one of the principal
  aims of this study is to demonstrate the application of the finite
  field methods to bulk electrolyte solutions, we are content with
  limiting ourselves to relatively small simulation cells for the sake
  of computational efficiency. While we believe our simulations to be
  sufficient for our current purposes, we have not investigated the
  potential effects of finite system size.} The ion-ion and ion-water
interactions were described with a Lennard-Jones potential and point
charges, using the parameters derived by Joung and Cheatham
\cite{joung2008determination}. Each simulation was approximately
50\,ns.

For all simulations we used the \textsmaller{LAMMPS} simulation
package \cite{plimpton1995sjc}. The particle-particle particle-mesh
Ewald method was used to account for long-ranged interactions
\cite{HockneyEastwood1988sjc}, with parameters chosen such that the
root mean square error in the forces were a factor $10^{5}$ smaller
than the force between two unit charges separated by a distance of
1.0\,\AA \cite{kolafa1992cutoff}. Dynamics were propagated using a
velocity-Verlet algorithm with a time step of either 1\,fs or
2\,fs. The temperature was maintained using a Nos\`{e}-Hoover
thermostat \cite{shinoda2004rapid,tuckerman2006liouville} at $T =
298$\,K. The geometry of the water molecules was maintained using the
\textsmaller{RATTLE} algorithm \cite{andersen1983rattle}. The
constant-$\mbf{E}$ Hamiltonian is implemented as standard in
\textsmaller{LAMMPS}. On the other hand, the constant-$\mbf{D}$
Hamiltonian was implemented `in-house'.\footnote{The source code is
  freely available at \url{https://github.com/uccasco/FiniteFields}.}

\section{Simulation Results}
\label{sec:sim}

\subsection{Electrolyte response to constant $\mbf{E}$}
\label{subsec:sim-E}

\begin{figure}[tb]
  \includegraphics[width=7.68cm]{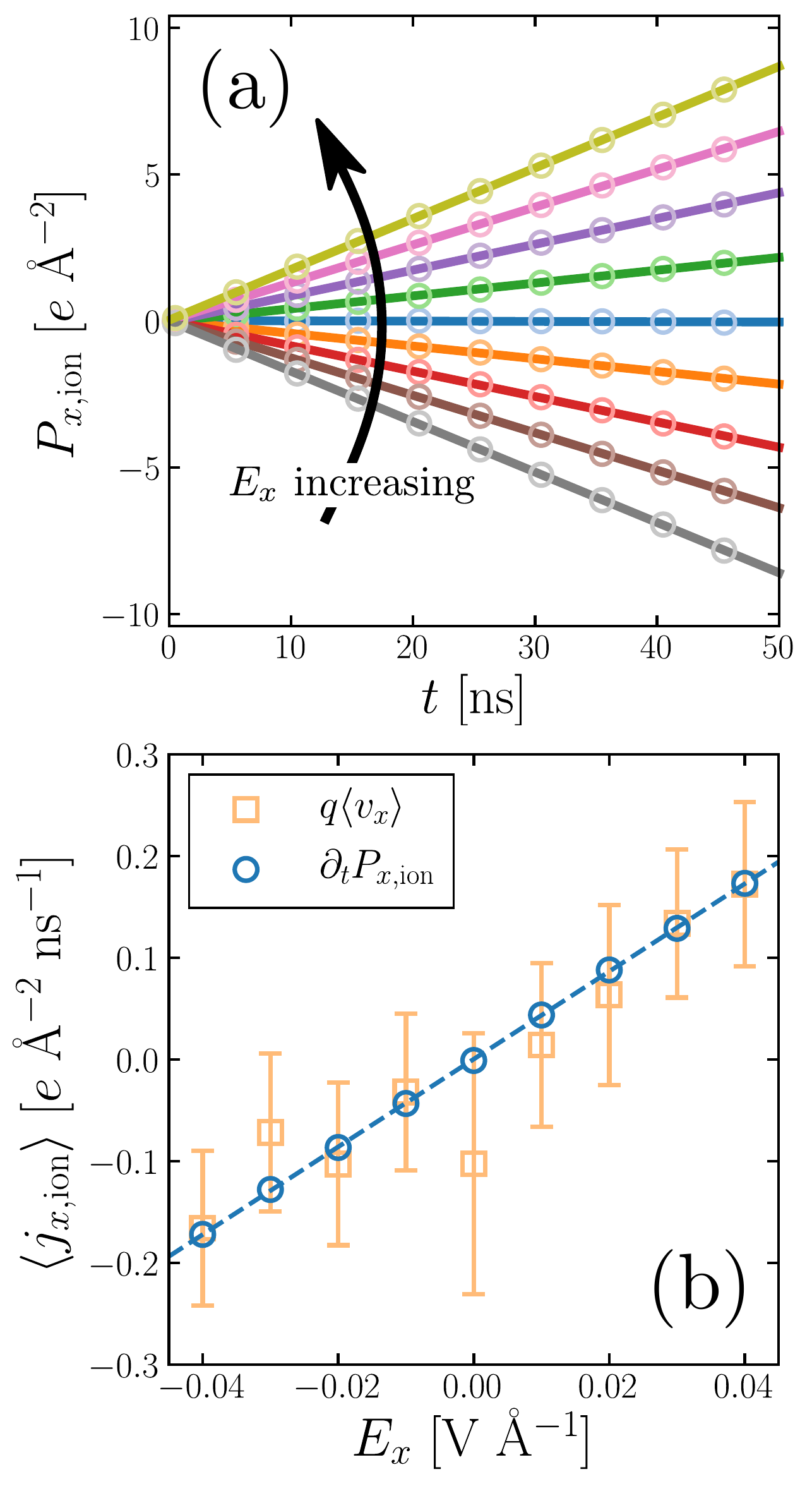}
  \caption{Ionic response of a 1\,M NaCl solution to a finite
    $\mbf{E}$ field. (a) $P_{x,\rm ion}$ vs $t$ for different
    $E_x$. Symbols show raw data from the simulations (only every
    100$^{\rm th}$ data point is shown for clarity), while solid lines
    show linear fits. The slope of each line gives the average ionic
    current density $\langle j_{x,\rm ion}\rangle = \langle J_{x,\rm
      ion}\rangle/\Omega$ for that field strength. (b) $\langle
    j_{x,\rm ion}\rangle$ vs $E_x$ obtained from the time evolution of
    $P_{x,\rm ion}$ (blue circles). The error estimate is smaller than
    the size of the symbols. The dashed line shows a linear fit; it is
    evident that for the range of $E_x$ used, we are in a linear
    response regime. The orange squares show $\langle j_{x,\rm ion}
    \rangle$ obtained from Eq.~\ref{eqn:j-vel}, which exhibits a far
    higher degree of noise.}
  \label{fig:E-response-1M}
\end{figure}

In Sec.~\ref{subsec:LR-sigma}, we presented a straightforward
derivation of the LR relation for $\sigma_{\rm ion}(0)$ within the
finite field formalism. In this section, we make use of the fact that
$\mathcal{H}_{\mbf{E}}$ is a fully microscopic Hamiltonian, and
explicitly simulate \ce{NaCl} solutions at finite $\mbf{E}$. As
mentioned previously, if an atom $i$ has a charge $q_i$, this amounts
to simply applying a force $\mbf{f}_{\mbf{E}} = q_i\mbf{E}$ to that
atom. Although this is what one might guess naively, applications of
this approach to bulk electrolyte solutions are surprisingly scarce,
although there are a number of examples in the biophysical literature
for calculating ionic fluxes through membranes (see
e.g. Refs.~\onlinecite{roux2004theoretical,modi2012computational,maffeo2012modeling}). In
Fig.~\ref{fig:E-response-1M}\,(a) we show the time evolution of
$P_{x,\rm ion}$, the $x$ component of ionic polarization, for 1\,M
\ce{NaCl}. As the itinerant polarization is the time integral of the
current density, we infer from Fig.~\ref{fig:E-response-1M}\,(a) that
we have reached a non-equilibrium steady state for each value of
$E_x$. Having obtained the time evolution of $P_{x,\rm ion}$, it is
straightforward to obtain $j_{x,\rm ion}$---whose dependence on $E_x$
is shown in Fig.~\ref{fig:E-response-1M}\,(b)---by linear
regression. For the range of $E_x$ studied, we see that the response
is remarkably linear. Moreover, the data appear essentially free from
noise, which is to be contrasted with the estimate of the current
density from,
\begin{equation}
  \label{eqn:j-vel}
  \langle j_{x,\rm ion} \rangle = \frac{1}{\Omega}\sum_{i}^{N_{\rm ion}} q_i\langle v_{x,i}\rangle,
\end{equation}
where $v_{x,i}$ is the $x$-component of the $i^{\rm th}$ ion's
velocity, which is also shown in
Fig.~\ref{fig:E-response-1M}\,(b). Although the general agreement with
the estimate based on $\partial_t P_{x,\rm ion}$ is sound, the degree of
noise is far higher. This is not unexpected, as we are effectively
attempting to extract the drift velocities imparted on the ions by the
field. The $v_{x,i}$ that enter the average in Eq.~\ref{eqn:j-vel} are
instantaneous velocities, and are thus distributed according to the
Maxwell-Boltzmann distribution. This gives rise to a relatively large
error on each individual measurement. Conversely, we measure $P_{x,\rm
  ion}$ at regular time intervals that are long (e.g. every 100\,ps)
compared to typical velocity autocorrelation times. This effectively
averages out the Maxwell-Boltzmann distribution, and greatly reduces
the error in the estimate of the drift velocity, and therefore also
the current density.

\begin{figure}[tb]
  \includegraphics[width=7.65cm]{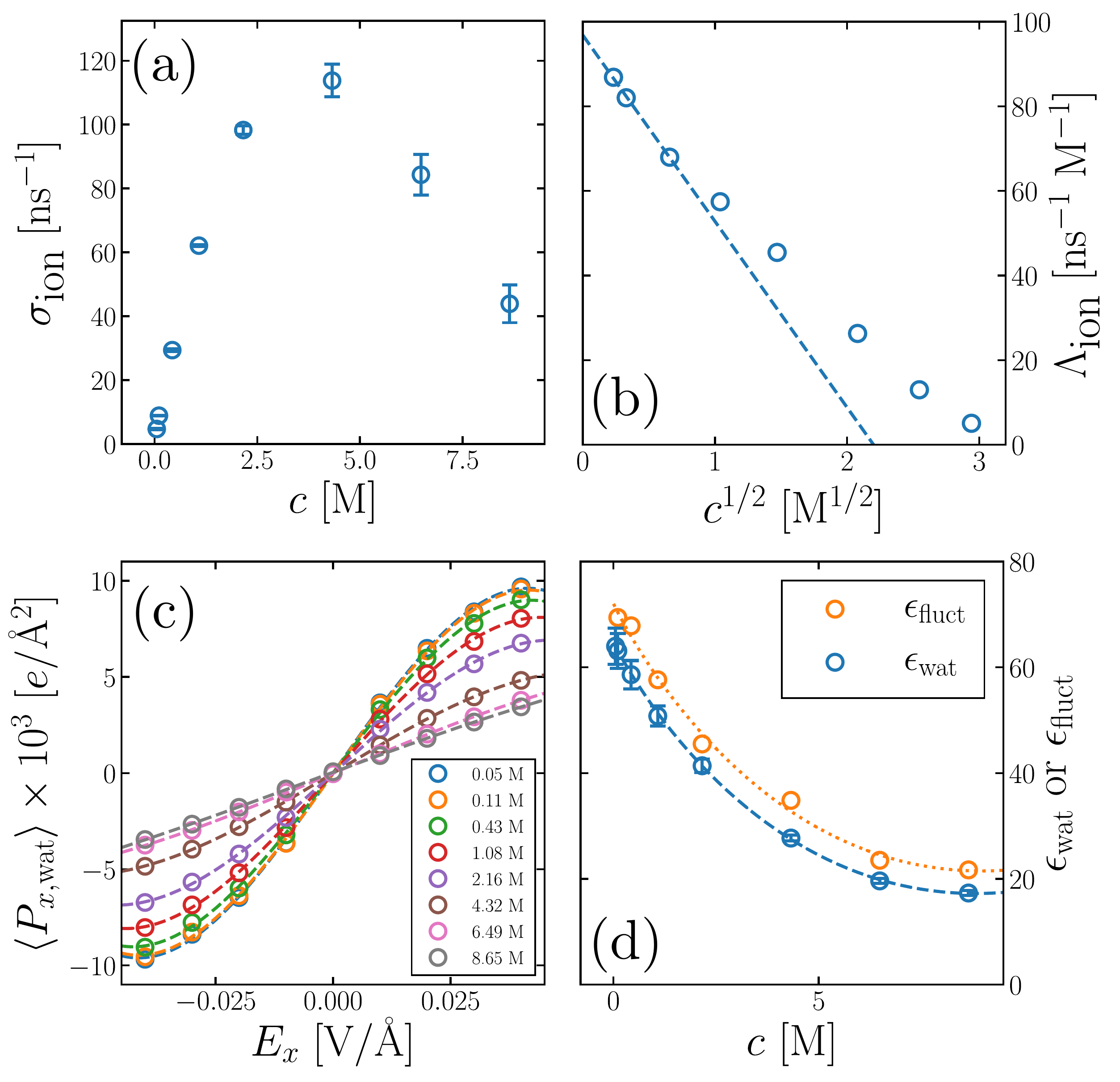}
  \caption{(a) $\sigma_{\rm ion}$ vs $c$. A maximum is observed at
    $c\approx 4$\,M. (b) $\Lambda_{\rm ion}$ vs
    $c^{1/2}$. Kohlrausch's law is obeyed at low concentrations. (c)
    $\langle P_{x,\rm wat} \rangle$ vs $E_x$ for different $c$ (see
    legend). The water responds non-linearly and also shows a $c$
    dependence. The dashed lines show fits to third order
    polynomials. (d) $\epsilon_{\rm wat}$ and $\epsilon_{\rm fluct}$
    vs $c$. Both measures of the dielectric constant are well
    described by the form $\epsilon = \epsilon_0 - Ac + Bc^{3/2}$,
    although $\epsilon_{\rm wat}$ (dashed line) is systematically
    lower than $\epsilon_{\rm fluct}$ (dotted line). Data for
    $c=0.05$\,M has been omitted due to inadequate statistics [panel
      (d) only].}
  \label{fig:E-response-full}
\end{figure}

The clear linear response of $\langle j_{x,\rm ion}\rangle$ seen in
Fig.~\ref{fig:E-response-1M} makes it simple to calculate $\sigma_{\rm
  ion}(0)$, whose dependence on $c$ is shown in
Fig.~\ref{fig:E-response-full}\,(a). It is interesting to observe that
$\sigma_{\rm ion}(0)$ exhibits a maximum in the conductivity at $c
\approx 4$\,M. In Fig.~\ref{fig:E-response-full}\,(b) we show the
molar ionic conductivities $\Lambda_{\rm ion} = \sigma_{\rm
  ion}/c$. According to Kohlrausch's law, for low concentrations
$\Lambda_{\rm ion}$ behaves as,
\begin{equation}
  \label{eqn:Lambda_ion}
  \Lambda_{\rm ion} = \Lambda_{\rm ion}^{(0)} - \mathcal{K}c^{1/2},
\end{equation}
where $\Lambda_{\rm ion}^{(0)}$ is the limiting value of $\Lambda_{\rm
  ion}$ i.e., the molar conductivity at infinite
dilution. $\mathcal{K}$ is a system-dependent constant, which accounts
for both electrophoretic and relaxation effects that impede the ionic
motion \cite{BokrisReddy1}. Although we only have limited data at low
$c$, fitting Eq.~\ref{eqn:Lambda_ion} for $c\lesssim 0.4$\,M and
extrapolating $c\to 0$ gives $\Lambda_{\rm ion}^{(0)} = 97\pm
2$\,ns$^{-1}$\,M$^{-1}$. This is to be contrasted with the
experimental value of $\Lambda_{\rm ion}^{(0)} =
114$\,ns$^{-1}$\,M$^{-1}$ \cite{KayeAndLaby}. Given the simple point
charge models used and the limited data at low $c$, this is a
remarkably satisfactory level of agreement.

In addition to the ionic conductivities, our simulations also allow us
to directly measure $\epsilon_{\rm wat}$ via the constitutive relation
Eq.~\ref{eqn:Pwat-constit}. In Fig.~\ref{fig:E-response-full}\,(c),
$\langle\mbf{P}_{\rm wat}\rangle$ vs $\mbf{E}$ is shown for all
concentrations studied. As well as exhibiting a concentration
dependence, the response is also noticeably non-linear. Nevertheless,
the simulation data are well approximated by a third order polynomial,
$\langle P_{x,\rm wat}\rangle = \langle P_{x,\rm wat}\rangle_0 +
\chi_{\rm eff}^{(1)}E_x + \chi_{\rm eff}^{(2)}E_x^2 + \chi_{\rm
  eff}^{(3)}E_x^3$, and we can extract the solvent dielectric constant
from the linear coefficient $\chi_{\rm eff}^{(1)}$,
\begin{equation}
  \chi_{\rm eff}^{(1)} = \frac{\epsilon_{\rm wat}-1}{4\pi}
\end{equation}
The concentration dependence of $\epsilon_{\rm wat}$ obtained in this
manner is shown in Fig.~\ref{fig:E-response-full}\,(d). Also shown is
the static dielectric constant obtained from the fluctuations of the
total solvent dipole moment,
\begin{equation}
  \epsilon_{\rm fluct}-1 = \frac{4\pi\beta}{\Omega}\langle(\delta M_{x,\rm wat})^2\rangle.
\end{equation}
Both $\epsilon_{\rm wat}$ and $\epsilon_{\rm fluct}$ depend on $c$ in
a similar fashion, and are well approximated by the commonly used form
\cite{friedman1982theory,FawcettLiquids},
\begin{equation}
  \label{eqn:eps_standard}
  \epsilon(c) = \epsilon_0 - Ac + Bc^{3/2}.
\end{equation}
However, it is clear that $\epsilon_{\rm fluct}$ is systematically
higher than $\epsilon_{\rm wat}$ across all concentrations. As
discussed in the introduction, this difference is a direct measure of
the dynamic contribution to the dielectric decrement
\cite{hubbard1977kinetic,hubbard1977dielectric,hubbard1979molecular,felderhof1984dielectric,chandra1993dielectric,chandra2000static}. As
shown by Caillol \etal\cite{caillol1986theoretical} the dynamic
contribution is due to a coupling between the ionic current and
solvent polarization,
\begin{equation}
  \label{eqn:DeltaEpsilon}
  \lim_{\omega\to 0}\epsilon_{\rm wat}(\omega) - \epsilon_{\rm fluct}
  = \frac{4\pi\beta}{3\Omega}\int_0^{\infty}\mrm{d}\tau\,\langle\delta\mbf{M}_{\rm wat}(\tau)\cdot\delta\mbf{J}_{\rm ion}(0)\rangle.
\end{equation}
Based on simulations between 1.8 and 2.2\,ns, and using comparable
system sizes to those in this article, Chandra concluded that the
dynamic contribution for aqueous \ce{NaCl} is finite but small:
Approximately two orders of magnitude smaller than the equilibrium
contribution \cite{chandra2000static}. To give a sense of the dynamic
contribution obtained in this work, in the limit $c\to 0$,
$\epsilon_{\rm fluct}$ is found to be $72\pm 3$, in excellent
agreement with previously computed values of the the dielectric
constant of SPC/E water
\cite{van1998systematic,aragones2010dielectric,zhang2014dipolar,braun2014transport,zhang2016computing1,zhang2016computing2}. In
contrast, the corresponding result for $\epsilon_{\rm wat}$ is $65\pm
1$, i.e. roughly 10\,\% smaller than $\epsilon_{\rm fluct}$.
Moreover, from Fig.~\ref{fig:E-response-full}\,(d), it appears that
the concentration dependence of the dynamic contribution is weak, and
thus becomes proportionally more significant at higher
concentrations. While our results are consistent with Chandra's
observation that the dynamic contribution is finite, we conjecture
that the differences in magnitude are due to difficulties in
converging the long time contributions to the time correlation
function in Eq.~\ref{eqn:DeltaEpsilon}, see Fig. \tcr{S9}. Our finding
of a larger dynamic contribution is also broadly consistent with Sega
\emph{et al} \cite{sega2014kinetic,sega2015kinetic}, who used a
nonequlibrium approach in which a fictitious field was applied only to
the ions. However, these authors also reported a dependence on the
force field used. We would like to stress that we have not attempted
to evaluate any potential effects of finite system size, and we have
extrapolated to infinite dilution from our relatively small simulation
cells. What our results demonstrate is an alternative approach to
investigating subtle effects such as the dynamic coupling between the
solvent and ions, based on a Hamiltonian for a system at constant
Maxwell field, $\mbf{E}$.

\subsection{Response to constant $\mbf{D}$}
\label{subsec:sim-D}

In contrast to its constant $\mbf{E}$ counterpart, the formulation of
the constant $\mbf{D}$ ensemble for finite temperature molecular
dynamics simulations is a more recent development. This statement,
however, warrants some qualification. In particular, if we set
$\mbf{D}=0$ in Eq.~\ref{eqn:H-D}, then the forces derived from
$\mathcal{H}_{\mbf{D}}$ take the same form as those presented by CLW
\cite{caillol1986theoretical,caillol1989electricalI,caillol1989electricalII}
for an electrolyte solution surrounded by a medium with dielectric
constant $\epsilon^\prime = 0$. Moreover, if we only set the
displacement field along one direction, say $D_z = 0$, and use tin
foil boundary conditions in the other two directions (so-called
`hybrid' boundary conditions \cite{zhang2016computing1}), then we
recover the popular Yeh-Berkowitz (YB) correction for simulations in a
slab geometry \cite{YehBerkowitz1999sjc}. Whereas in the YB scheme it
is necessary to introduce a vacuum region, no such constraints are
imposed on the system by $\mathcal{H}_{\mbf{D}}$, a fact that was
recently exploited in Refs.~\onlinecite{zhang2016finite} in the study
of electrolyte/solid interfaces. In this study, we have removed all
extended interfaces entirely.

\begin{figure}
  \includegraphics[width=7.65cm]{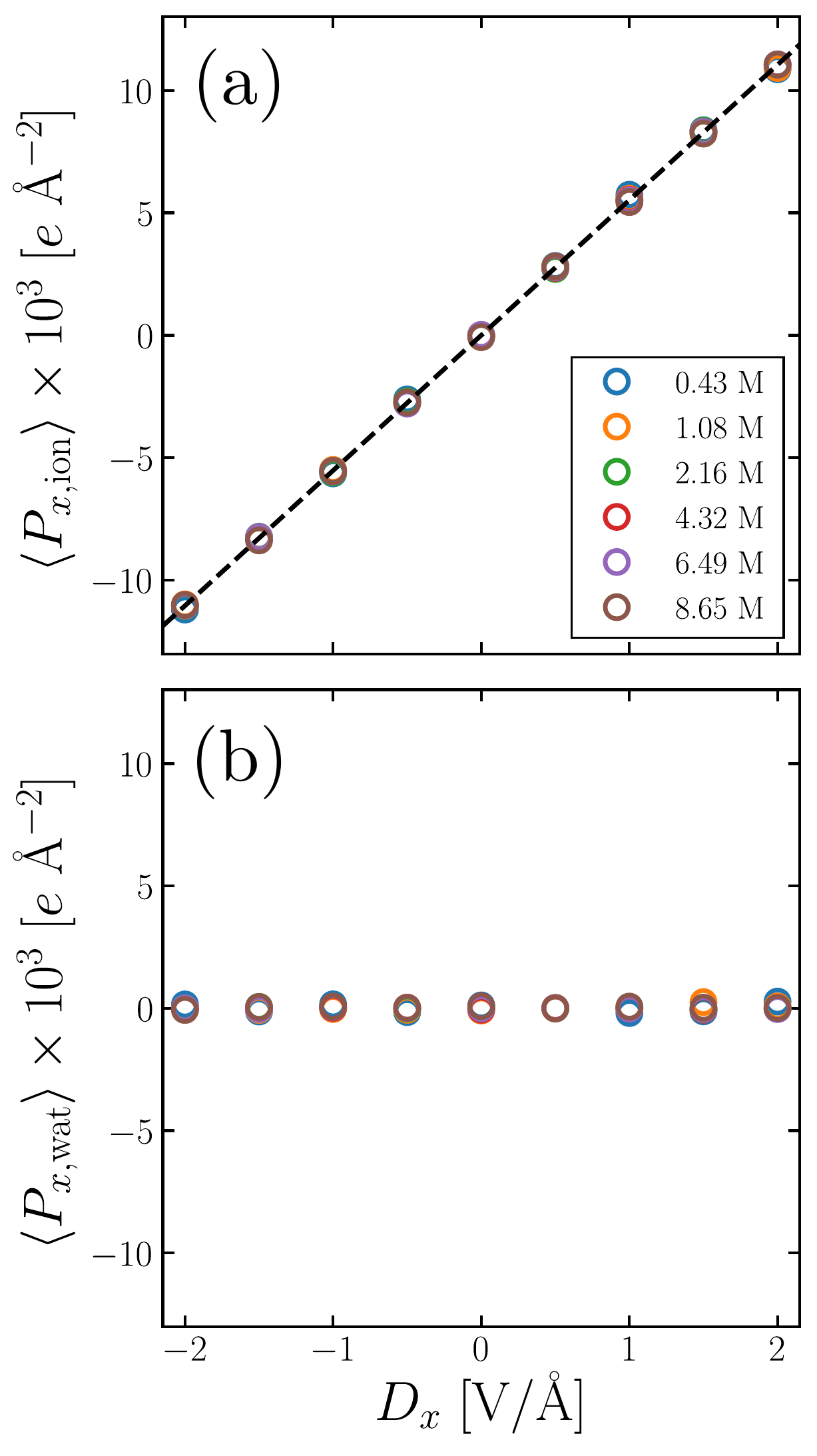}
  \caption{The ions completely screen the $\mbf{D}$ field. (a)
    $P_{x,\rm ion}$ vs $D_x$ for different $c$, as indicated by the
    legend. The polarization response from the ions is independent of
    $c$. The dashed line shows $P_{x,\rm ion} = D_x/4\pi$, the
    theoretical result for a conductor. (b) $P_{x,\rm wat}$ vs
    $D_x$. There is negligible solvent response. Error estimates are
    smaller than the size of the symbols. Data for $c\lesssim 0.4$\,M
    has been omitted due to inadequate statistics.}
  \label{fig:D-response}
\end{figure}

\begin{figure}[tb]
  \includegraphics[width=7.65cm]{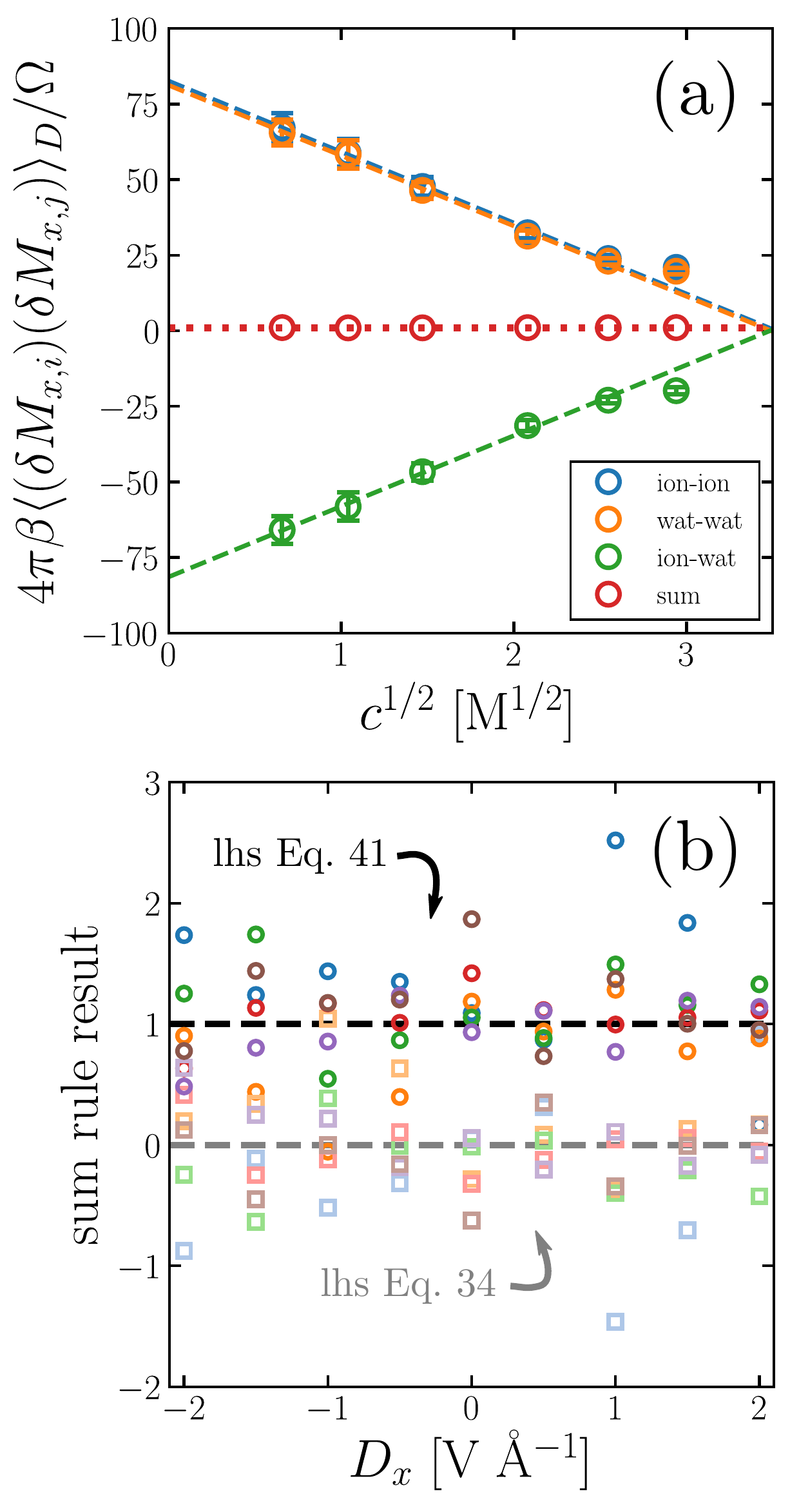}
  \caption{The Stillinger-Lovett conditions place strict requirements
    on the polarization fluctuations. (a) $4\pi\beta\langle(\delta
    M_{x,i})(\delta M_{x,j})\rangle$ vs $c^{1/2}$, with
    $i,j=\text{`ion' or `wat'}$. Both $\langle(\delta M_{x,\rm
      ion})^2\rangle$ (blue) and $\langle(\delta M_{x,\rm
      wat})^2\rangle$ (orange) decrease approximately linearly with
    $c^{1/2}$. Conversely, $\langle(\delta M_{x,\rm ion})(\delta
    M_{x,\rm wat})\rangle$ increases (green), such that the sum rule,
    Eq.~\ref{eqn:SL-full1}, is satisfied (red). (b) The left hand
    sides of Eq.~\ref{eqn:SL-wat} (squares) and Eq.~\ref{eqn:SL-ion}
    (circles); the black and gray dashed lines show the respective
    theoretical predictions. Different colors refer to different $c$,
    as in Fig.~\ref{fig:D-response}. Data for $c\lesssim 0.4$\,M has
    been omitted due to inadequate statistics.}
  \label{fig:sumrules}
\end{figure}

Despite the above similarities to the work of CLW and YB,
$\mathcal{H}_{\mbf{D}}$ has only recently been identified as the
Hamiltonian for finite temperature MD simulation in the constant
$\mbf{D}$ ensemble. It is therefore unsurprising that the response of
bulk electrolyte solutions to finite displacement fields has not been
widely studied. In Fig.~\ref{fig:D-response} we show $\langle P_{x,\rm
  ion}\rangle$ and $\langle P_{x,\rm wat}\rangle$ vs $D_x$ for
different concentrations. Clearly, all the response originates from
the ions, consistent with the discussion presented in
Sec.~\ref{subsec:SL-conditions}. The dashed line in
Fig.~\ref{fig:D-response}\,(a) shows the theoretical result for a
conductor, $4\pi\langle P_{x,\rm ion}\rangle = D_x$, to which the
simulation data conforms excellently. We therefore conclude, as
expected, that $\langle E_x\rangle = 0$ in our simulation, which was
the starting point for the derivation of the SL conditions (see
Sec.~\ref{subsec:SL-conditions}). Do the simulations also confirm the
quantitative theoretical predictions of the SL conditions given by
Eqs.~\ref{eqn:SL-wat} and~\ref{eqn:SL-ion}?  This is indeed the case,
as demonstrated in Fig.~\ref{fig:sumrules}. From
Fig.~\ref{fig:sumrules}\,(a), the expected anticorrelation of ionic
and water polarization fluctuations is observed. In both
Figs.~\ref{fig:D-response} and~\ref{fig:sumrules}\,(a), we have
omitted data for $c\lesssim 0.4$\,M, owing to insufficient statistics
for the lowest concentrations (see below). Thus while the polarization
fluctuations appear to decrease with $c^{1/2}$ in
Fig.~\ref{fig:sumrules}\,(a), we cannot preclude deviations from this
behavior at low concentration. Fig.~\ref{fig:sumrules}\,(b) shows the
left hand sides of Eqs.~\ref{eqn:SL-wat} and~\ref{eqn:SL-ion}, as
measured from simulation. They are clearly consistent with the
theoretical predictions; gathering statistics from all simulations
gives,
\begin{align}
  \frac{4\pi\beta}{\Omega}\bigg[\langle (\delta M_{x,\rm ion})^2\rangle + \langle (\delta M_{x,\rm ion})(\delta M_{x,\rm wat})\rangle\bigg]
  &= 1.07 \pm 0.11, \label{eqn:sim-sumrule-ion}\\[7pt] 
  \frac{4\pi\beta}{\Omega}\bigg[\langle (\delta M_{x,\rm wat})^2\rangle + \langle (\delta M_{x,\rm ion})(\delta M_{x,\rm wat})\rangle\bigg]
  &= -0.05 \pm 0.11. \label{eqn:sim-sumrule-wat}
\end{align}

Combined with the theoretical results of
Sec.~\ref{subsec:SL-conditions}, these simulation results are a
powerful demonstration that, in a bulk electrolyte, fluctuations in
the ionic and solvent polarization are inextricably linked.

As mentioned in the introduction, the finite field method for constant
$\mbf{D}$ was previously used by Pache and Schmidt to calculate
$\epsilon_{\rm wat}$ from the change in the polarization
\cite{pache2018molecular}. This was done by coupling the $\mbf{D}$
field only to the water polarization, which was motivated by the fact
that, although a transient water polarization was observed when
coupling the $\mbf{D}$ to the total polarization, only the ions
contributed upon reaching equilibrium. From the theoretical and
simulation results presented here, this can be understood as a
manifestation of the SL conditions. Applying the $\mbf{D}$ field only
to the water will affect the fluctuations and likely violate the SL
conditions. Developing optimal strategies for computing $\epsilon_{\rm
  wat}$ for electrolyte systems from constant $\mbf{D}$ simulations
requires further theoretical considerations that lie beyond the scope
of the current article.

\begin{figure}[tb]
  \includegraphics[width=7.65cm]{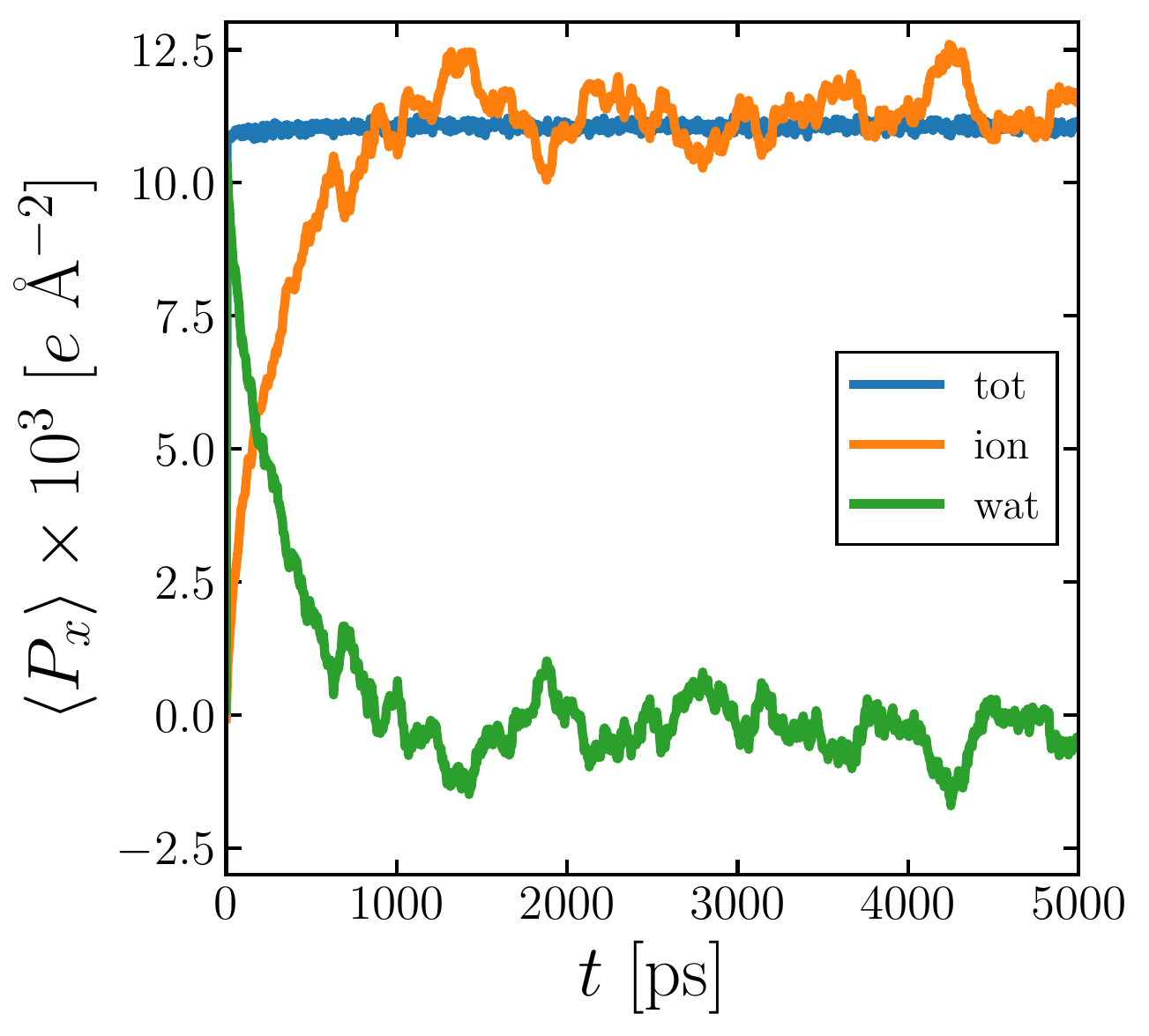}
  \caption{Time evolution of the total polarization $P_x$, and its
    contributions from the water and ions ($P_{x,\rm wat}$ and
    $P_{x,\rm ion}$, respectively), for $c\approx 0.11$\,M and $D_x =
    2.0$\,V/\AA. The initial configuration was taken from an
    equilibrated $\mbf{D}=\mbf{0}$ simulation. While $P_x$ attains its
    equilibrium value relatively quickly, $P_{x,\rm wat}$ and
    $P_{x,\rm ion}$ take much longer to relax (approx. 1\,ns), and
    also exhibit correlations over long timescales.}
  \label{fig:D-relax}
\end{figure}

We end this section with a comment regarding the timescales for
relaxation toward equilibrium. In Fig.~\ref{fig:D-relax} we show the
time evolution of the total polarization $P_x$, along with its
contributions $P_{x,\rm wat}$ and $P_{x,\rm ion}$, for $c\approx
0.11$\,M and $D_x = 2.0$\,V/\AA. It is clear that while $P_x$ attains
its equilibrium value relatively quickly, $P_{x,\rm wat}$ and
$P_{x,\rm ion}$ take on the order of 1\,ns to relax. It is also
apparent that there exists correlations over long timescales for
$P_{x,\rm wat}$ and $P_{x,\rm ion}$. Thus while $P_x$ may appear well
converged on short timescales, there is a real risk of inadequate
sampling of the equilibrium phase space distribution
function. Although such effects are exaggerated for low
concentrations, these results are potentially concerning for \emph{ab
  initio} MD studies of electrolyte systems at constant $\mbf{D}$, but
may guide future strategies for tackling such issues.

\section{Summary}
\label{sec:summary}

The purpose of this work was to investigate the behavior of bulk
aqueous electrolyte solutions using the finite field methods developed
in
Refs.~\onlinecite{zhang2016computing1,zhang2016computing2,zhang2016finite}. In
comparison to existing theoretical frameworks, this has offered great
conceptual simplifications. Using the Hamiltonian for constant Maxwell
field $\mbf{E}$, we derived the linear response formula for the ionic
conductivity without reference to the vector potential. The
particularly pleasing aspect of this approach is that the Hamiltonian
used to derive the linear response relation is the same as that used
to derive the forces for molecular dynamics simulations. This was put
into practice here to obtain an ionic conductivity at infinite
dilution within 15\,\% of experimental values. In addition, this
approach enabled us to extract the dielectric constant of the solvent
water from its response to finite $\mbf{E}$, which was seen to
decrease with increasing electrolyte concentration. We also observed
that the dielectric constant measured from the response was
systematically smaller than that estimated from fluctuations of the
solvent polarization, which can be taken as a direct measure of the
dynamic coupling between fluctuations of the solvent polarization and
ionic current.

We also used the finite field method for constant electric
displacement $\mbf{D}$ to derive the Stillinger-Lovett conditions that
relate ionic and solvent polarization fluctuations. In addition to
providing a mechanical picture with which to understand the SL
conditions, this approach avoids the need to relate the perturbing
field to the cavity field. We exploited this fact in our simulations
to explicitly measure the system's response to finite $\mbf{D}$, which
supported the theoretical predictions. At equilibrium, we found that
all polarization response emanates from the ions. We also observed
that relaxation of the ionic polarization to equilibrium could be a
slow process, especially for dilute solutions. The anticorrelations
imposed by the Stillinger-Lovett conditions, however, causes the
solvent polarization to relax in a manner such that the total
polarization appears to attain its equilibrium value on relatively
short timescales.

One of the major motivations for the development of the finite field
methods for finite temperature simulations was to mitigate spatial
finite size effects that lead to incomplete screening of charged
insulator/electrolyte interfaces. While we remain optimistic that such
techniques will prove a useful tool in \emph{ab initio} studies of
such systems, our results emphasize the need to exercise caution with
respect to proper sampling of the equilibrium phase space distribution
function. Our results also showcase the application of different
electrostatic boundary conditions to electrolyte systems beyond the
slab geometry employed for solid-liquid interfaces. The finite field
methods can therefore be viewed as an additional tool with which to
study electrolyte systems, and may find uses in e.g. the computation
of dielectric spectra
\cite{schroder2008collective,schroder2008computation,sega2013calculation,sega2014kinetic,rinne2014dissecting,sega2015kinetic},
or ion transport through membranes
\cite{roux2004theoretical,modi2012computational,maffeo2012modeling}. The
key challenge faced now is to generalize this approach to systems with
applied electric fields and polarization varying in space. In other
words, how to define polarization density for the ions for which
recourse to the multipole expansion is not possible?

\section{Supplementary Material}

See supplementary material for results obtained with concentration
dependent simulation cell sizes such that the pressure remained
constant, results obtained with hybrid boundary conditions, and a
comparison to the Green-Kubo approach.

\begin{acknowledgments}
  Peter Wirnsberger is thanked for many helpful discussions,
  especially regarding implementation of the constant-$\mbf{D}$
  ensemble in \textsmaller{LAMMPS}. Thomas Sayer is thanked for his
  comments on a draft of this manuscript. We are grateful for
  computational support from the UK Materials and Molecular Modelling
  Hub, which is partially funded by EPSRC (EP/P020194), for which
  access was obtained via the UKCP consortium and funded by EPSRC
  grant ref EP/P022561/1. S.J.C. is supported by a Royal Commission
  for the Exhibition of 1851 Research Fellowship.
\end{acknowledgments}

\bibliography{../../cox}

\end{document}